\documentclass{aa}  

\usepackage{graphicx}
\usepackage{array}
\usepackage{tabularx}

\usepackage{txfonts}
\usepackage{hyperref}
\hypersetup{
    colorlinks=true,
    linkcolor=blue,
    filecolor=magenta,      
    urlcolor=cyan,
    citecolor=blue,
    }

\def\ltsima{$\; \buildrel < \over \sim \;$}
\def\simlt{\lower.5ex\hbox{\ltsima}}
\def\gtsima{$\; \buildrel > \over \sim \;$}
\def\simgt{\lower.5ex\hbox{\gtsima}}

\def\msun{{\rm M_{\odot}}}

\def\be{\begin{equation}}
\def\ee{\end{equation}}

\def\del#1{{}}
\def\ltsima{$\; \buildrel < \over \sim \;$}
\def\simlt{\lower.5ex\hbox{\ltsima}}
\def\gtsima{$\; \buildrel > \over \sim \;$}
\def\simgt{\lower.5ex\hbox{\gtsima}}

\usepackage{xcolor}
\usepackage{comment}

\newcolumntype{L}{>{$}l<{$}}
\newcolumntype{C}{>{$}c<{$}}
\newcolumntype{R}{>{$}r<{$}}

\usepackage{booktabs}
\usepackage{siunitx}

\begin{document}

   \title{Active galactic nucleus outflows accelerate when they escape the bulge}
   \titlerunning{AGN outflow acceleration}

   \author{K. Zubovas\orcid{0000-0002-9656-6281}\inst{1} \and M. Tartėnas\orcid{0009-0006-7373-180X}\inst{1}}

   \institute{Center for Physical Sciences and Technology, Saulėtekio al. 3, Vilnius LT-10257, Lithuania\\
              \email{kastytis.zubovas@ftmc.lt}
             }

   \date{Received ...; accepted ...}

 
  \abstract
   {Large-scale outflows driven by active galactic nuclei (AGNs) are an important element of galaxy evolution. In addition, detailed analysis of their properties allows us to probe the activity history of the galactic nucleus and, potentially, other properties of the host galaxy. A recent paper presents detailed radial velocity profiles of outflows in ten AGN host galaxies and shows a common trend of approximately constant velocity in the centre followed by rapid acceleration outside $R_{\rm tr} \sim 1-3$~kpc.}
   {We wish to understand whether the observed acceleration is a consequence of the AGN-driven outflows clearing the gaseous bulges of the host galaxies and beginning to expand into a region of negligible gas density.}
   {We used the 1D semi-analytical code {\sc Magnofit} to calculate outflow propagation in each of the ten galaxies, assuming a constant AGN luminosity and an isothermal bulge density profile, and leaving the gas fraction and total mass of the bulge as free parameters. We also considered the effect of different gas density profiles, variations in bulge velocity dispersion, AGN luminosity, and the effect of outflow fragmentation.}
   {Our simplest model can fit six outflow profiles essentially perfectly, while another can be fit if the bulge gas density profile is shallower than isothermal. A shallower density profile also improves the fit in the central regions of the remaining three outflows, but they accelerate faster than our models predict in the outskirts. We propose that this could be evidence of significant gas cooling and star formation that reduce the total mass of outflowing gas.}
   {We show that a simple AGN-driven wind feedback model can explain the detailed velocity profiles of real outflows in local AGN hosts. The free parameters of our model have values that fall well within reasonable ranges. This suggests that the simple scenario we envisioned is close to the true conditions governing the general trends of large-scale outflow expansion.}


   \maketitle
%

\section{Introduction}

It has now been almost two decades since the first detections of massive fast outflows in galaxies, often driven by active galactic nuclei \citep[AGNs; e.g.][]{Morganti2007AA, Tremonti2007ApJ, Veilleux2009ApJS, Feruglio2010AA, Sturm2011ApJ, Rupke2011ApJ}. The global properties of these outflows, such as their velocity, mass flow rate, total momentum, and kinetic power, correlate well with the AGN luminosity \citep{Cicone2014AA, Gonzalez2017ApJ, Fluetsch2019MNRAS, Lutz2020AA}, albeit with large scatter \citep{Marasco2020AA}. They provide a means for the AGN to communicate a significant fraction of its luminosity to the host galaxy interstellar medium (ISM); the resultant AGN feedback is a necessary ingredient in galaxy formation models, establishing the observed correlations between the supermassive black hole (SMBH) mass and galaxy properties \citep[e.g.][]{King2003ApJ, King2010MNRASa} and the shape of the high-mass end of the galaxy mass function \citep[e.g.][]{Croton2006MNRAS, Bower2006MNRAS, Vogelsberger2014MNRAS, Schaye2015MNRAS, Tremmel2019MNRAS, Nelson2019MNRAS}. Furthermore, outflows act as dynamical footprints of past AGN luminosity variations and so can help us understand the activity history of their host galaxies \citep{Nardini2018MNRAS, Zubovas2020MNRAS, Zubovas2022MNRAS}.

Given their importance, it has become clear that spatially resolved observations of AGN-driven outflows are needed to advance the field of research, for multiple reasons. First, they reveal the multi-phase nature of the outflows \citep{Fluetsch2021MNRAS}, which has been identified as one of the essential gaps in our knowledge \citep{Cicone2018NatAs}. They also act as a proxy for tracking the temporal variations in AGN luminosity, since outflowing gas at different distances from the nucleus is presumably powered by the AGN output emitted at different times \citep{Zubovas2020MNRAS}. Finally, radial profiles of outflow properties can help distinguish between various theoretical models proposed to explain their driving mechanism and their impact on the host galaxies.

Of the many theoretical models that attempt to explain AGN-driven outflow properties and their impact on their host galaxy, the wind-driven outflow model stands out \citep[see e.g.][]{King2015ARAA}. Within this framework, the AGN luminosity drives a quasi-relativistic wind from the accretion disc. The wind carries a kinetic power ($\dot{E}_{\rm w}$) of $\sim 0.05L_{\rm AGN}$ at a velocity ($v_{\rm w})$ of $\sim 0.1$c. It shocks the relatively static ISM and gets heated to a temperature ($T_{\rm sh}$) of $\sim 10^{10}$~K. At this temperature, the only potentially relevant cooling process is inverse Compton (IC) scattering against the AGN radiation field. If it is efficient, in the sense that the shocked wind energy is radiated away faster than it can be transferred adiabatically to the surrounding ISM, only the momentum of the wind is transferred to the ISM, leading to a momentum-conserving (or momentum-driven) outflow \citep{King2003ApJ, King2010MNRASa}. Conversely, if cooling is inefficient, all of the wind energy is transferred to the ISM, leading to an energy-driven outflow \citep{King2005ApJ, Zubovas2012ApJ}. Where and under what conditions the two situations occur is a matter of some debate: if the shocked wind is a single-temperature plasma, we can expect cooling to be efficient in the central several hundred parsecs \citep{King2003ApJ} and, thus, the two types of outflows to be clearly separated radially. On the other hand, if the plasma is two-temperature, cooling becomes much less efficient \citep{Faucher2012MNRASb} and an energy-driven wind exists everywhere; however, in an inhomogeneous ISM, the outflow expands in directions of least resistance, potentially leaving cold gas clumps behind and exposed only to the momentum of the wind \citep{Zubovas2014MNRASb}. As mentioned above, spatially resolved outflow observations can help distinguish between the two scenarios.

\citet[][hereafter M25]{Marconcini2025NatAs} present detailed 3D models of ionised gas outflows in ten nearby active galaxies. They find that in seven galaxies, the outflows accelerate significantly, with their velocity increasing by a factor of $>2$, starting at $1-3$~kpc from the nucleus. In three other galaxies, the observations only covered distances up to $\sim 0.7$~kpc from the nucleus; nevertheless, there are hints of outflow acceleration in the outer parts of the probed regions as well. All ten galaxies exhibit approximately constant or slightly decreasing velocities from the centre to the point where acceleration begins. The authors interpret this remarkable trend as evidence of the outflow transitioning from a momentum-driven to an energy-driven phase once it expands beyond a critical radius that does not strongly depend on galaxy properties.

We explored a different possibility: that the outflows are energy-driven throughout the galaxy but accelerate once they have cleared the bulge of the host galaxy. The radius at which the outflows begin to accelerate is comparable to the typical size of spiral galaxy bulges \citep{Kim2016ApJS}; it appears natural to expect the outflow to accelerate once it breaks out of the relatively gas-rich region into a gas-poor halo. Using the 1D outflow simulation code {\sc Magnofit}, we calculated the radial velocity profiles of outflows in each galaxy analysed in M25, assuming a purely energy-driven outflow propagating through a bulge with an isothermal density distribution and a finite extent. The base model used only two free parameters --- the gas fraction and total mass of the bulge. Our simple model can fit six observed outflow profiles almost perfectly, while the other four can be made to fit via relatively straightforward adjustments of model assumptions. In all cases, the values chosen for the free parameters agree well with observational constraints.

The paper is structured as follows. We begin by presenting the {\sc Magnofit} code and the analytical expectations of energy-driven outflow propagation in Sect. \ref{sec:method}. Then we present the results of our base model in Sect. \ref{sec:results}. We consider the ways to improve the fit in Sect. \ref{sec:variations}, the effect of AGN luminosity uncertainties in Sect. \ref{sec:present_day_agn}, and the effect of changing the bulge mass and velocity dispersion in Sect. \ref{sec:varsigma}. In Sect. \ref{sec:spatial_structure} we present some predictions regarding the spatial structure of outflows that may be unveiled with future observations. Finally, in Sect. \ref{sec:momentum_criticism} we highlight some of the drawbacks of the theoretical interpretation for outflow acceleration given in M25. We summarise and conclude in Sect. \ref{sec:sum}.

\section{Method} \label{sec:method}

\begin{table*}
\centering
\caption{Galaxy properties.}
\label{tab:galaxy_properties}
\begin{tabular}{lcccccccc}
\hline
ID & $M_{\mathrm{BH}}$ & $\sigma$ & $\langle v\rangle_{\rm norm}$  & $M_{\mathrm{bulge}}$ & $M_{\mathrm{bulge}}$ & $f_{\rm g}$ & $M_{\mathrm{bulge}}$ & $f_{\rm g}$ \\
   & ($10^6 \,M_\odot$) & (km s$^{-1}$) & (km s$^{-1}$) & ($10^9 \,M_\odot$) & ($10^9 \,M_\odot$) & (\%) & ($10^9 \,M_\odot$) & (\%) \\
   & (from M25)        & (from M25)& ($<1$~kpc) & from S19 & (fit value) & (fit value) & (flatter fit) & (flatter fit)\\
\hline
NGC 7582   & $55$  & 140  & 595  & $14.0$ & $16$  & 9.3 & -&-\\
IC 5063    & $174$ & 175  & 2040 & $35.3$ & $42$  & 0.52 & -&-\\
Circinus   & $3.5$ & 80 & 491$^*$  & $1.5$  & $1.2$ & 3.4 & -&-\\
Centaurus A& $59$  & 140  & 546$^*$  & $14.8$ & $2.5$ & 12 & -&-\\
NGC 4945   & $1.4$ & 100 & 1060$^*$ & $0.729$& $2.1$ & 0.089 & -&-\\
IC 1657    & $47$  & 143  & 765  & $12.3$ & $27$  & 3.7 & -&-\\
NGC 1365   & $20$  & 114  & 482  & $6.17$ & $6.8$  & 9.3 & $5.5$ & 14\\
NGC 2992   & $23$  & 130  & 705  & $6.9$  & $6.8$ & 2.8 & $12$ & 14\\
NGC 5643   & $2.8$ & 80 & 670$^*$  & $1.25$ & $1.4$ & 1.1 & $1.5$ & 2.5\\
NGC 5728   & $34$  & 125  & 690  & $9.46$ & $6.6$ & 4.8 & $9.0$ & 21\\
\hline
\end{tabular}
\newline
\begin{minipage}{\textwidth}
\vspace{0.2cm}
\small
\textbf{Notes:} Columns show, from left to right, the galaxy ID, black hole mass (from Table 1 of M25), bulge velocity dispersion (Supplementary Fig. 3 of M25 and Cosimo Marconcini, private communication), mean outflow velocity within the central region used to normalise the velocity profile (Cosimo Marconcini, private communication), bulge mass derived from $M_{\rm BH}$ using the relation in \citet{Schutte2019ApJ}, bulge mass selected to produce the best fit to the outflow velocity profile, and gas fraction $f_{\rm g}$ selected in the same way. For the last four galaxies, the final two columns show the bulge mass and gas fraction required when fitting a flatter bulge gas density profile. \\
$^*$ - in these galaxies, the outflow begins to accelerate within the central kiloparsec, so we recalculated the mean velocity including only the points interior to the acceleration region.
\end{minipage}
\end{table*}

We calculated the properties of expanding outflows using {\sc Magnofit}\footnote{Freely available here: \url{https://github.com/zadrras/magnofit}}, a 1D code purpose-built for such applications \citep{Zubovas2022MNRAS}. It numerically integrates the equation of motion of a spherically symmetric outflow driven by the energy of the AGN wind, assuming full adiabaticity, in a generic background gravitational potential and gas density distribution. The only constraints are that both the potential-generating mass distribution ($M_{\rm pot}$) and the gas distribution ($M$), as well as their first and second radial derivatives, can be expressed analytically. The equation of motion is \citep[see Appendix A of][for the derivation]{Zubovas2022MNRAS}
\begin{equation} \label{eq:eom}
    \dddot{R} = \frac{2}{MR} \left(\frac{\eta}{2}L_{\rm AGN} - A\right) - B,
\end{equation}
where $R$ is the radius of the contact discontinuity, $\eta \sim 0.1$ is the radiative efficiency of accretion, $L_{\rm AGN}$ is the AGN luminosity and the terms $A$ and $B$ are, respectively,
\begin{equation}\label{eq:term_A}
    A = \dot{M}\dot{R}^2 + M\dot{R}\ddot{R} + \frac{2G \dot{R}}{R^2}M\left(M_{\rm pot} + \frac{M}{2}\right)
\end{equation}
and
\begin{equation}\label{eq:term_B}
  \begin{split}
    B &= \frac{\ddot{M} \dot{R}}{M} + \frac{\dot{M} \dot{R}^2}{M R} + \frac{2\dot{M} \ddot{R}}{M} + \frac{\dot{R} \ddot{R}}{R} + \\ &
    +\frac{G}{R^2}\left[M_{\rm pot}\frac{\dot{M}}{M} + \dot{M} + \dot{M}_{\rm pot}
       - \left(M_{\rm pot}+\frac{M}{2}\right)\frac{\dot{R}}{R}\right].
  \end{split}
\end{equation}
The time derivatives of mass terms are defined as $\dot{M}\equiv \dot{R}\partial M/\partial R$ and $\ddot{M} \equiv \ddot{R}\partial M/\partial R + \dot{R}^2 \partial^2 M/\partial R^2$.

In a strongly idealised case, the equation can be solved analytically. When both $M$ and $M_{\rm pot}$ follow an isothermal distribution --- $M_{\rm pot} = 2 \sigma^2 R/G$ (equivalently, $\rho = \sigma^2/\left(2\pi G R^2\right)$) and $M = f_{\rm g} M_{\rm pot}$ --- and the AGN luminosity ($L_{\rm AGN} = lL_{\rm Edd} = 4\pi G M_{\rm BH} c l/\kappa$) is constant in time, the outflow quickly attains a constant velocity,
\begin{equation} \label{eq:ven}
    \dot{R} \simeq \left(\frac{2 \pi \eta G^2 M_{\rm BH} c l}{3 f_{\rm g} \kappa \sigma^2}\right)^{1/3} \simeq 394 \left(\frac{M_{\rm BH,7} l}{f_{0.1}\sigma_{150}^2}\right)^{1/3} \, {\rm km\, s}^{-1},
\end{equation}
where in the second equality we scale $M_{\rm BH}$ to $10^7\,\msun$, $f_{\rm g}$ to 0.1 and $\sigma$ to 150~km~s$^{-1}$. When the outflow clears the bulge, the surrounding gas density becomes negligible and the outflowing gas mass stays constant. This allows the outflow to accelerate, but an analytical solution is impossible.

{\sc Magnofit} uses several parameters to generate the outflow properties. The galaxy is composed of two extended components, called the `bulge' and the `halo'. Each has a total mass, characteristic radius, density profile and gas fraction, which can be set independently or constrained using observationally derived relations between them and the SMBH mass. The sum of gas masses of the two components within the outflow radius determines the outflowing gas mass, while the rest of the mass comprises the potential-generating mass in Eqs. \ref{eq:eom}-\ref{eq:term_B}. We aimed to keep our models as simple as possible, so we built them starting with the observationally constrained SMBH mass, $M_{\rm BH}$, taken from Table 1 of M25, and velocity dispersion $\sigma$ derived from stellar continuum fitting (see Supplementary Fig. 3 of M25). We used the relation given in Eq. 8 of \citet{Bandara2009ApJ} to determine the virial mass of the galaxy. This value falls in the range from $\sim 5\times 10^{11} \, \msun$ for NGC 4945 to $\sim 10^{13} \, \msun$ for IC 5063; however, its precise value had negligible influence on our results, since we are interested in the very central regions of the galaxy, where the bulge component dominates completely. Similarly, the halo mass (equal to the virial mass minus the bulge mass), virial radius, concentration and density profile also do not affect the results, as long as they fall within reasonable value ranges. We used the default relationships in {\sc Magnofit}, without the uncertainties, to determine these values, and set the gas fraction in the halo to $10^{-3}$. We chose an isothermal bulge density profile extending from the centre to the outer radius of the bulge, $R_{\rm b}$, which is determined by $\sigma$ and the bulge mass $M_{\rm bulge}$: $R_{\rm b} = GM_{\rm bulge}/\left(2\sigma^2\right)$. We considered two options for choosing $M_{\rm bulge}$. In the first, we took its value from the observationally constrained $M_{\rm BH} - M_{\rm bulge}$ relation as given in \citet{Schutte2019ApJ}. In the second, we left it as a free parameter to be determined by fitting the observed outflow velocity profile. The gas fraction in the bulge, $f_{\rm g}$, was kept as a free parameter. Finally, we assumed the AGN keeps shining at a constant luminosity $L_{\rm AGN} = 0.5L_{\rm Edd}$ throughout the outflow lifetime. This luminosity is much higher than the present-day luminosity of the investigated galaxies; we return to this point in Sect. \ref{sec:present_day_agn}. We summarise all the relevant parameters and fitting results in Table \ref{tab:galaxy_properties}.

With these parameter choices, we ran the calculation of outflow expansion for each of the ten galaxies, aiming to produce a close fit to the observed outflow velocity profile. We divided the observed profile into two regions: central and outer. We defined the central region as either the central kiloparsec (as in M25 when calculating the normalised velocities) or the region before the outflow starts accelerating, whichever is smaller. In the central region, the outflow velocity does not depend on $M_{\rm bulge}$, so we could choose the value of $f_{\rm g}$ to produce an exact match to the average observed velocity in the region. Once we fixed the best-fitting value of $f_{\rm g}$, we varied $M_{\rm bulge}$ until the accelerating portion of the outflow matched the observations as closely as possible, by minimising the value of $\chi^2 \equiv \sum_{\rm r>R_{\rm b}} (v_{\rm obs} - v_{\rm model})^2 / v_{\rm obs}^2$. In cases where the observed velocity profile could not be fully matched, we chose the value of $M_{\rm bulge}$ that allowed the model outflow velocity to intersect with the point of the highest observed velocity. Although this does not provide the lowest value of $\chi^2$, we believe it is important to determine the conditions for the outflow to accelerate to the maximum observed velocity, as we discuss in Sect. \ref{sec:variations}.

\section{Results} \label{sec:results}

\begin{figure*}
\includegraphics[width=\textwidth]{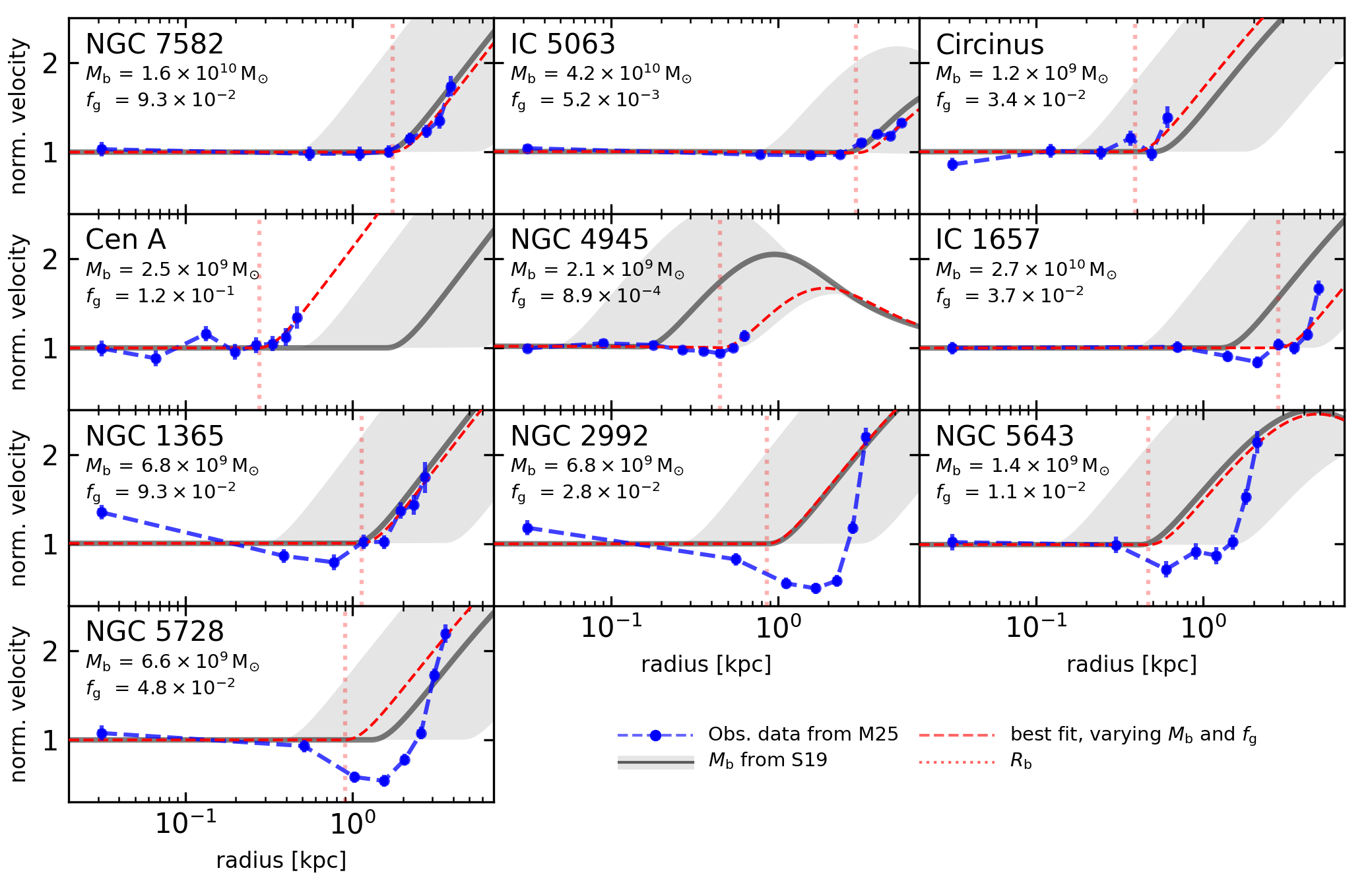} 
\caption{Observed (blue points) and modelled outflow velocity profiles for all ten galaxies. In each galaxy, all velocities are normalised to the mean observed outflow velocity in the central kiloparsec or within the modelled bulge radius, whichever is lower. Panels are arranged approximately from best to worst match. The black line and shaded region correspond to the $M_{\rm bulge}$ taken from the observed $M_{\rm BH}-M_{\rm bulge}$ relation and its $\pm 0.68$~dex scatter \citep{Schutte2019ApJ} when the value of $f_{\rm g}$ is chosen to fit the velocity in the central region. The dashed red line is the result of choosing both $f_{\rm g}$ and the value of $M_{\rm bulge}$ to produce the best fit to the observed outflow profile. The vertical dotted red line shows the bulge radius given by the best-fit value of $M_{\rm bulge}$, which is also the point where the outflow begins to accelerate.}
\label{fig:all_fit}
\end{figure*}

\begin{figure}
\includegraphics[width=\columnwidth, trim={0 1cm 0 0}, clip]{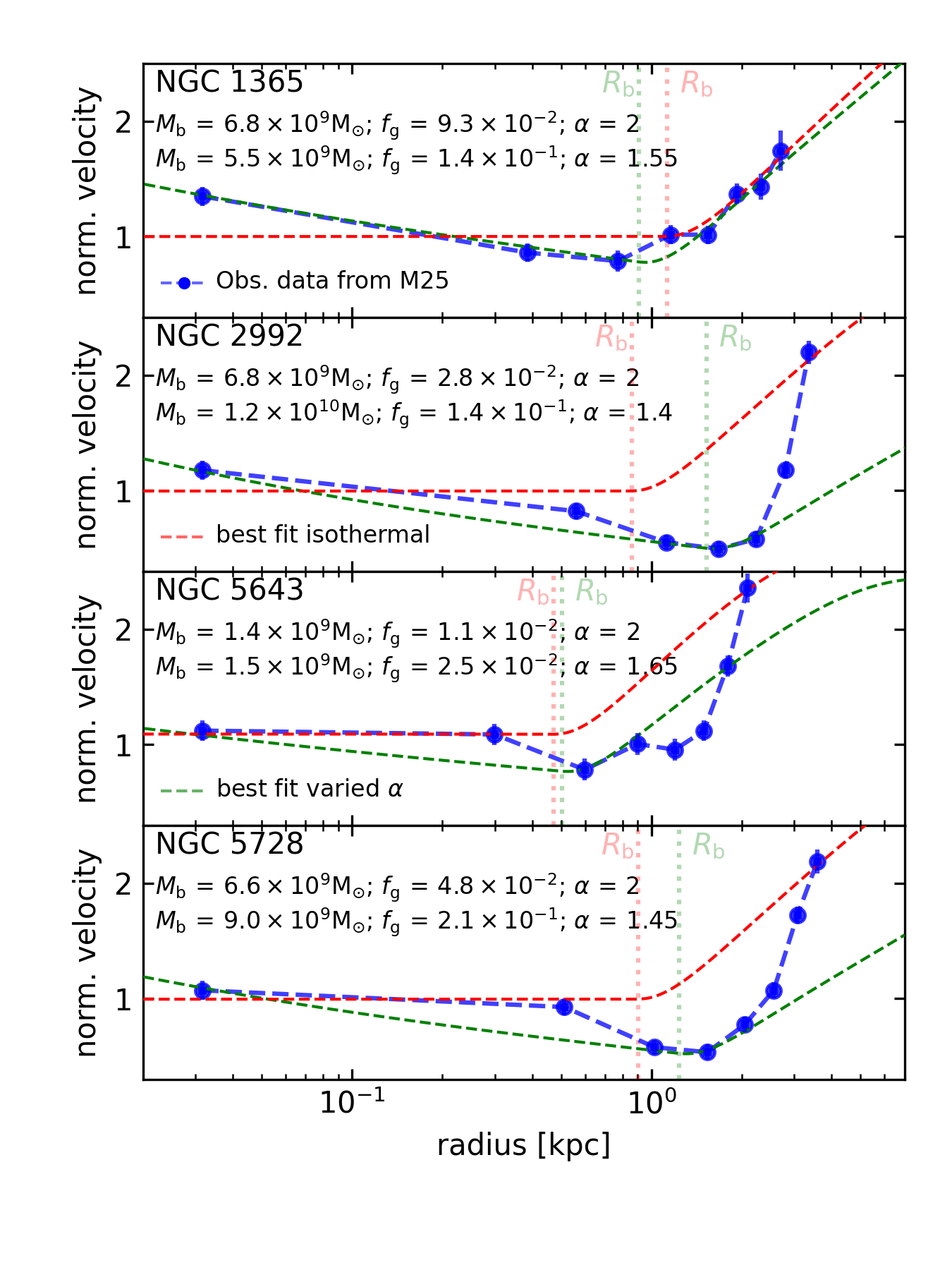} 
\caption{Effect of varying the slope of the gas density distribution. Top panel: Observed outflow in NGC 1365 (blue points) compared with our fiducial model (dashed red line; same as in Fig. \ref{fig:all_fit}) and a model where the density profile is $\rho \propto R^{-1.55}$ (green dashed line). Other panels: Same as top panel but for NGC 2992, NGC 5643, and NGC 5728, from top to bottom. In these three cases, we chose the values of the free parameters to match the innermost observed point and the point with the lowest velocity (see the main text for our motivation). In each panel, we give the values of $M_{\rm bulge}$, $f_{\rm g}$, and the slope exponent, $\alpha,$ for each model.}
\label{fig:varslope}
\end{figure}

\begin{figure}
\includegraphics[width=\columnwidth]{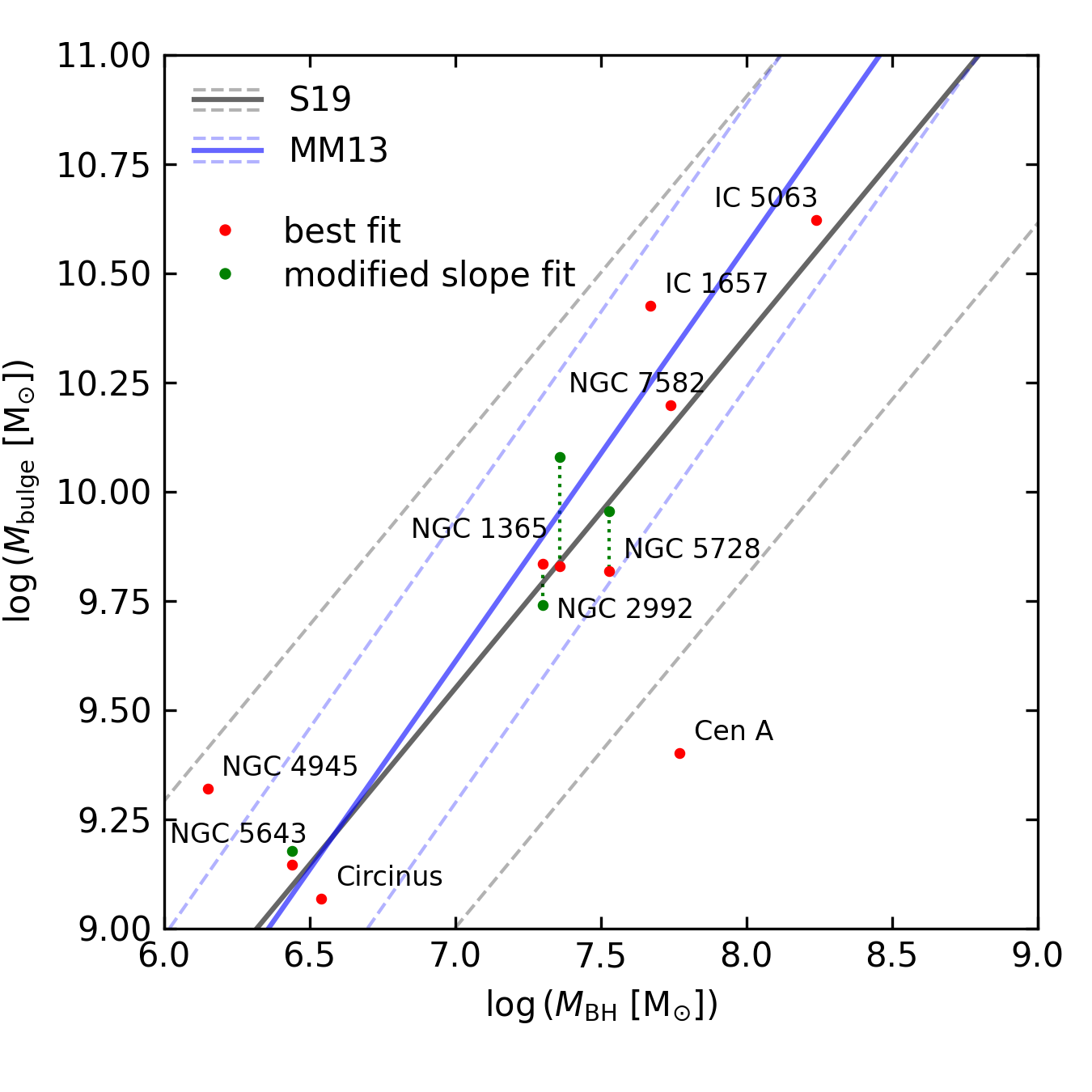} 
\caption{Black hole and bulge masses of our best-fit base models (red points), plotted against SMBH masses. Green points show the bulge masses in the fits with different bulge density profile slopes. The grey line shows the observed $M_{\rm BH} - M_{\rm bulge}$ relation from \citet{Schutte2019ApJ}, with dashed lines representing the $\pm 0.68$~dex scatter. The blue line shows the equivalent relation from \citet{McConnell2013ApJ}, with dashed lines representing the $\pm 0.34$~dex scatter.}
\label{fig:mbh_mbulge}
\end{figure}

In Fig. \ref{fig:all_fit} we show the results of our modelling for all ten galaxies, sorted from best to worst match. Our model fits the data very well in at least a few cases (e.g. NGC 7582 and IC 5063). In Centaurus A, the fit is very good, although the required bulge mass and radius differ significantly from observational expectation and from observational estimates \citep[$R_{\rm bulge} \sim 5.5$~kpc, cf.][]{Sahakyan2013ApJ}; however, we are interested only in the gaseous component of the bulge, which may be significantly smaller than the stellar bulge, especially in early-type galaxies. Alternatively, the observed upturn in outflow velocity may be spurious and the actual acceleration may happen much further out, where currently available observations do not probe due to their limited field of view. 

In one galaxy, NGC 1365, our model fits the outer part rather well, but cannot fit the decreasing velocity in the central region. The same is true for three more galaxies --- NGC 2992, NGC 5643, and NGC 5728; additionally, here our model fails to fit the outer part as well, predicting much more gradual acceleration than what is observed. A rather simple extension to the model can significantly improve the fit to the central regions of the four galaxies, as we show in Fig. \ref{fig:varslope}. There, we show result of fitting a model where we relaxed the assumption that the bulge density profile is isothermal. This leads to one more fit parameter, the density profile slope $\alpha$ (where in the isothermal case, $\alpha = 2$). For NGC 1365, we performed fitting in much the same way as for the isothermal models, first varying $\alpha$ and $f_{\rm g}$ to fit the innermost three points, then adjusting $M_{\rm bulge}$ to minimise $\chi^2$ of the other five. The fit is clearly much better; note that it required a slightly less massive bulge with a slightly higher gas fraction. 

For the other three galaxies, we were only interested in fitting the innermost point and the lowest-velocity point, while we chose $M_{\rm bulge}$ to allow the outflow to start accelerating just outside the lowest-velocity point. In these cases, the bulge mass is slightly higher and so is the gas fraction. The slope of the accelerating part of the outflow is quite similar between the isothermal and flatter-slope cases in all four galaxies. This is expected, because in that region, the AGN is pushing out the whole mass of the bulge gas and is fully outside the potential-generating non-gaseous bulge mass distribution, so only the total mass of the bulge, and its gas, are relevant to the outflow dynamics. We treat the two models as extreme cases, with the isothermal one fitting the highest-velocity point and the flatter-slope one fitting the slowest point, and discuss the possible reasons for the rapid acceleration in Sect. \ref{sec:variations}.

In most of our models, both isothermal and flatter-slope, the best-fit bulge mass is very similar to the one given by the observed relation as showcased in \ref{fig:mbh_mbulge}. Nine out of ten galaxies have bulge masses that fall within the $\pm 0.68$~dex scatter of the \citet{Schutte2019ApJ} relation, while eight have masses falling within the narrower $\pm 0.34$~dex scatter of the \citet{McConnell2013ApJ} relation. Only Centaurus A, discussed above, has a best-fit bulge mass significantly lower than either relation predicts. Furthermore, NGC 4945 has a bulge mass that is significantly higher than predicted by \citet{McConnell2013ApJ}; however, we note that their sample used to derive this relation is mostly comprised of galaxies with $M_{\rm bulge} > 10^{10} \, \msun$.

The gas fractions required in our simulations span a range between $f_{\rm g}<10^{-3}$ and $f_{\rm g}=0.1$; they reach $f_{\rm g} = 0.21$ in the flatter-slope models. We define gas fraction as the ratio of gas mass to the total mass of the bulge. Observational studies sometimes use a different definition, where the gas fraction is the ratio of gas mass to stellar mass. This corresponds to\footnote{We investigate the central regions of galaxies, where the dark matter contribution is expected to be negligible, so we think it is justified to neglect it in these calculations.} $M_{\rm g}/M_* \approx f_{\rm g}/(1-f_{\rm g})$; since the values of $f_{\rm g} \ll 1$, the ratio is very similar to $f_{\rm g}$. This range aligns well with the gas fractions observed in spiral galaxies at low redshift \citep{Combes2013AA, Fisher2013ApJ, Diaz-Garcia2020AA}. Our models typically have somewhat lower gas fractions than observed galaxies, which is to be expected, since bulges are often gas-poor compared to discs \citep{Fisher2013ApJ}. The total gas masses, $M_{\rm g} = f_{\rm g} M_{\rm bulge}$, range between a few times $10^6 \, \msun$ in NGC 4945 to over $10^9 \, \msun$ in NGC 7582; such values are compatible with estimates of gas content in the inner regions of spiral galaxies \citep{Boker2003AA}. In all cases, they are lower than the total gas mass given in Table 1 of M25, although the value for NGC 7582 is very close. We conclude that both bulge masses and gas fractions, and the resulting bulge gas masses, are reasonable.

\section{Discussion} \label{sec:discuss}

\subsection{Rapid outflow acceleration in outer regions} \label{sec:variations}

Three galaxies --- NGC 2992, NGC 5643, and NGC 5728 --- have outflows that accelerate much faster than our models predict. A change in the slope of the density profile improves the fit to the velocity profile in the central region, but this comes at the expense of making the fit to the outer parts of the profile even worse. Although using different model parameters can fit both the lowest and highest observed outflow velocities, fitting them simultaneously with our simple model is impossible. We propose that this rapid acceleration may be caused by a continuous decrease in total outflowing mass, probably caused by star formation within the outflow.

Star-forming regions have been discovered in many AGN outflows \citep{Maiolino2017Natur, Gallagher2019MNRAS}. This discovery validates earlier analytical and numerical predictions regarding gas cooling in massive outflows \citep{Zubovas2013MNRAS, Zubovas2014MNRASa, Richings2018MNRAS, Richings2018MNRASb}. While the outflow is moving through the gaseous bulge, star formation does not affect its dynamics very much: Eqs. 18 and 21 in \citet{Zubovas2014MNRASa} show that the star formation rate in an outflow should always be much lower than the rate at which it entrains new material. However, once the bulge is cleared, no new material joins the outflow. At that point, cooling and star formation can lead to a potentially significant reduction in outflowing mass. The cooling time estimated in the same paper and by \citet{Richings2018MNRASb} is generally shorter than the outflow dynamical time, so the ionised outflow component may lose a significant fraction of its gas before it expands to more than a few times the bulge radius. The gas fraction estimated by our base models may correspond to this leftover mass, rather than the total mass of gas that was pushed out of the bulge. Our models for two of the three galaxies, NGC 2992 and NGC 5643, predict bulge gas masses that are $<0.01$ times the value of $M_{\rm gas}$ given in M25, so it is certainly possible that their full outflows are more massive than our fiducial estimates. The flatter-slope models that we used to fit the lowest-velocity points in the observed profiles all require higher gas fractions and bulge masses, leading to total gas masses that are $2.4-8.8$ times higher than in the base models. Such masses are still well below the $M_{\rm gas}$ values in NGC 2992 and NGC 5643, and close to the $M_{\rm gas}$ value of NGC 5728, indicating that it is indeed possible that the galaxy had more substantial amounts of gas while the outflow was clearing the bulge.

Observations of all three galaxies show possible evidence of positive AGN feedback (NGC 2992: \citealt{Xu2024ApJ}; NGC 5643: \citealt{Cresci2015AA}; NGC 5728: \citealt{Shin2019ApJ}). However, this evidence is limited to the central kiloparsec and to the interaction between the ionised outflow and the pre-existing gas clouds, rather than fragmentation of the outflowing gas itself. Nevertheless, this possibility should be explored with future observations. In particular, if it was discovered that significant star formation occurs within the outflows in these three galaxies but not in the ones without such rapid outflow acceleration, it would significantly support our explanation.

In addition to acceleration, some of our models, for example those for NGC 4945 and NGC 5643, show the outflow decelerating at an even greater distance. This happens because as the outflow expands, the gravitational potential becomes progressively more dominated by the halo component. In a Navarro–Frenk–White (NFW) halo, the inner part within the scale radius has an approximate density slope $\rho \propto R^{-1}$, which leads to an energy-conserving outflow decelerating. When the outflow moves past the scale radius, which is $\sim 25-100$~kpc in our models, it begins to accelerate once again. We predict that observations sensitive to outflows at distances of up to $10$~kpc from the nucleus should reveal some deceleration.

\subsection{Present-day AGN luminosity} \label{sec:present_day_agn}

\begin{figure}
\includegraphics[width=\columnwidth]{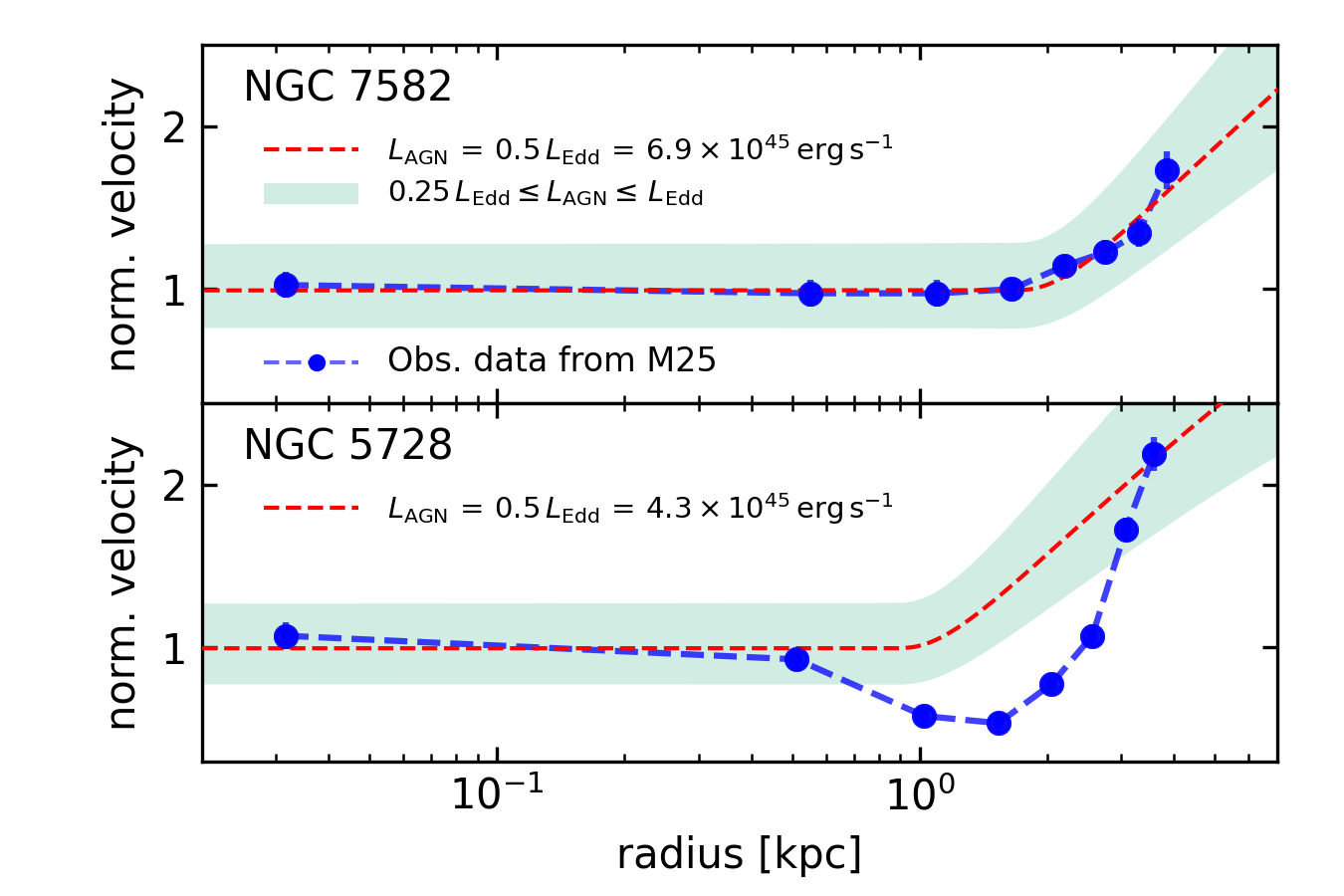} 
\caption{Effect of varying $L_{\rm AGN}$ on outflow velocity profiles, while keeping all other parameters fixed. Blue points are observational data from M25 and the dashed red line is our best-fit model (as in Fig. \ref{fig:all_fit}) using $L_{\rm AGN} = 0.5 L_{\rm Edd}$. The shaded region represents the effect of changing the AGN luminosity by a factor of 4, between $0.25 L_{\rm Edd}$ and $L_{\rm Edd}$.}
\label{fig:varlagn}
\end{figure}

All ten galaxies we analysed are AGN hosts, but most have Eddington ratios much lower than we used in our models, from $l \approx 0.0027$ in Centaurus A \citep{Ogawa2021ApJ} to $l \approx 0.235$ in NGC 5643 \citep{Garcia-Bernete2021AA}. We chose a high value of $l = 0.5$ instead of using the actual observed luminosity for two reasons. First of all, the present-day luminosities of some galaxies are below $0.01 L_{\rm Edd}$, which is the approximate limit at which accretion flow transitions between thin disc and radiatively inefficient or advection-dominated accretion flow (RIAF, ADAF respectively) modes \citep{Best2012MNRAS, Sadowski2013MNRAS, Koudmani2024MNRAS}. The physical model of AGN wind-driven outflows relies on thin disc radiation to launch the wind, so within this paradigm, these galaxies would be unable to drive any outflows at their present-day luminosity. This does not invalidate the whole model, because the AGN could have been much brighter even in the relatively recent past. Both statistical \citep{Schawinski2015MNRAS} and theoretical arguments \citep{King2015MNRAS} suggest that typical AGN episodes last $\sim 10^5$~yr; given that the age of the large-scale outflows is $> 1$~Myr, there has been ample time for the luminosity to have decayed while the outflow still exists \citep{King2011MNRAS, Zubovas2023MNRAS}.

Secondly, the precise value of AGN luminosity during driving is not strictly constrained, because it is tightly related to the gas density. The analytical expression for the outflow velocity (Eq. \ref{eq:ven}) shows that $v_{\rm out} \propto \left(l/f_{\rm g}\right)^{1/3}$, i.e. a reduction in AGN luminosity by some factor is equivalent to increasing the gas fraction by the same factor. This is the case in our modelled outflows in the central (constant-velocity) region and remains almost true in the accelerating region. We show the absolute effect of AGN luminosity variation in Fig. \ref{fig:varlagn}. Higher AGN luminosity results in higher outflow velocity in the central region, by up to a factor of $2^{1/3} \approx 1.26$, and lower luminosity results in a corresponding velocity decrease. 
Outside the bulge, the outflow accelerates slightly faster when the luminosity is higher, but the effect is mild: at the radius where the best fit model reaches a normalised velocity $v_{\rm out}/v_{\rm norm} = 2$, the maximum-luminosity model shows a velocity increase by a factor of $\sim2.06$ from its value in the central region; for the lowest-luminosity model, the corresponding factor is $\sim1.97$. The effect of adjusting the gas fraction by a factor of 2 is identical to changing the AGN luminosity, as expected from the analytical estimate. As a result, our model results can be seen as providing an upper limit to the value of gas fraction in the galaxy bulges; if the AGN luminosity was always similar to the present-day value, the required bulge gas fraction would be a factor of $2-50$ lower than our estimates.

\subsection{Influence of the bulge mass and velocity dispersion}\label{sec:varsigma}

\begin{figure}
\includegraphics[width=\columnwidth]{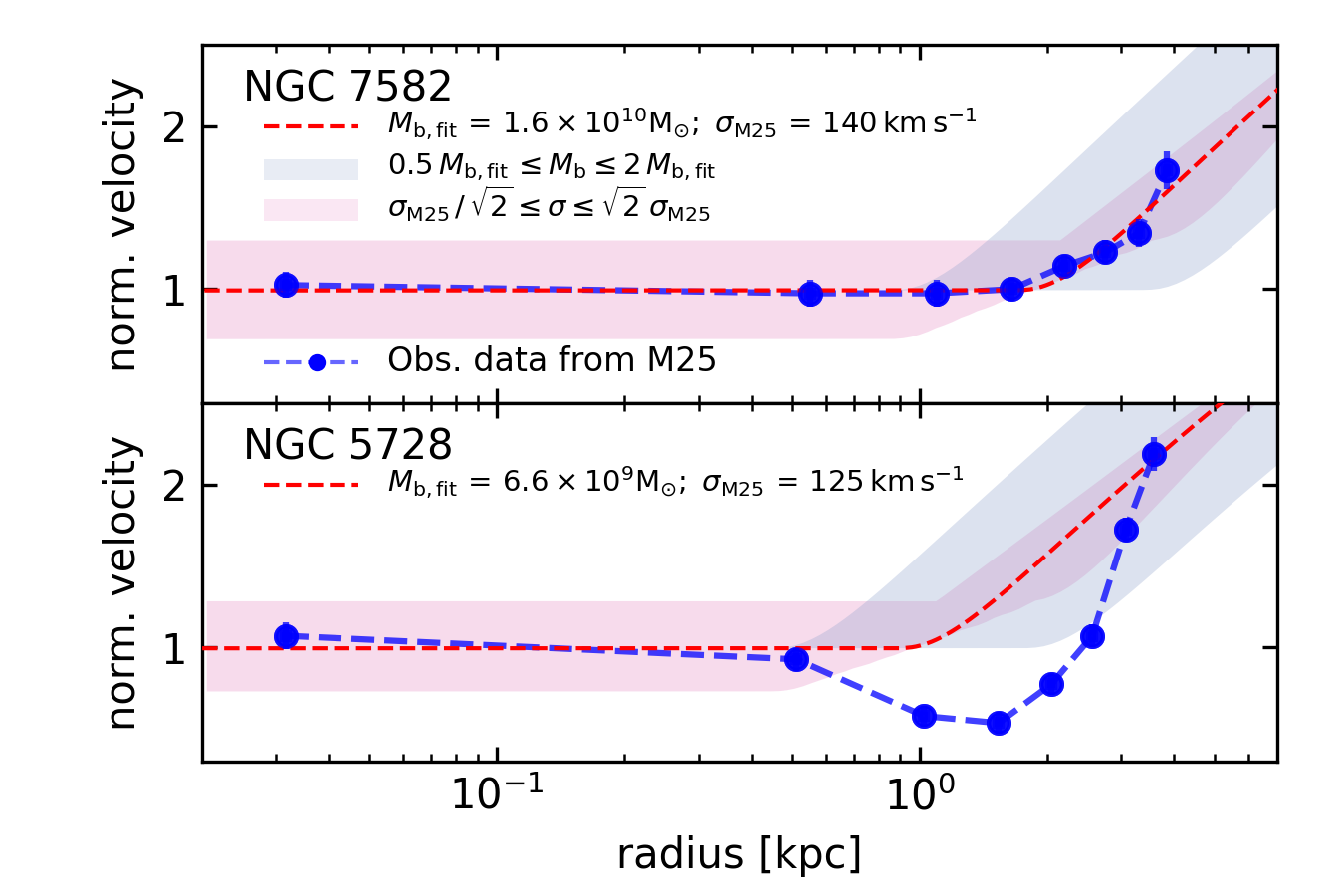} 
\caption{Same as Fig. \ref{fig:varlagn} but showing the effect of varying the bulge mass (grey shading) and velocity dispersion (purple shading). The bulge mass varies by a factor of 4 around the fiducial value, and the velocity dispersion by a factor of 2.}
\label{fig:varsigma}
\end{figure}

In Fig. \ref{fig:varsigma} we show the effect on our model velocity profiles of varying the bulge mass and velocity dispersion, while keeping all other parameters the same. Bulge mass variations have no effect on outflow velocity in the central region. This is expected, because a different bulge mass simply results in a change in the bulge radius but not in the density inside the bulge, which is the parameter governing the initial outflow velocity. Lighter bulges allow earlier outflow acceleration, since $R_{\rm b} \propto M_{\rm b}$, while more massive bulges keep the outflow at constant velocity for longer. Variations of bulge velocity dispersion have an effect on all regions of the outflow: higher $\sigma$ leads to lower initial velocity, since $v_{\rm out} \propto \sigma^{-2/3}$, but earlier acceleration, since $R_{\rm b} \propto \sigma^{-2}$. Both effects can be compensated for by a change in $f_{\rm g}$ and $M_{\rm b}$.

\subsection{Spatial structure of outflows} \label{sec:spatial_structure}

An adiabatic, energy-driven outflow should have approximately constant kinetic energy rate at each radius. Since $\dot{E}_{\rm kin} = \dot{M}_{\rm out} v^2/2$, the observed velocity profile translates directly to a prediction regarding the mass outflow rate. We expect a constant $\dot{M}_{\rm out}$ in the central regions of galaxies, which begins to decrease as the outflow accelerates. This expectation agrees very well with results of idealised hydrodynamical simulations \citep{Zubovas2024AA}. Determination of the mass flow rates in the real galaxies will be a powerful test of our model.

Notably, the mass flow rate prediction should only apply to the ionised gas outflow, which was the focus of M25. Outflows are generally multi-phase \citep[e.g.][]{Fluetsch2021MNRAS}, so we expect that the ionised outflows in these ten galaxies expand together with many denser and colder clumps. The shocked AGN wind flows around pre-existing dense clumps, which are left exposed only to the wind momentum (\citealt{Zubovas2014MNRASb, Tartenas2025sub}). As a result, we expect that a significant fraction of the cold gas is outflowing at lower velocities, especially in the central region. In the outskirts, cold gas can only exist if it condenses from the cooling outflowing gas \citep{Zubovas2014MNRASa}. Its velocity at the moment of condensation should be similar to that of the ionised gas but later the dense gas should lag behind as it is no longer pushed so efficiently. So the cold gas at each radius should consist of clumps condensing at that radius moving with a velocity similar to the ionised gas and clumps that condensed closer to the nucleus but moved further away, moving slower than the ionised gas. This will result in a velocity gradient for dense gas at each radius; this is another testable prediction as more spatially resolved multi-phase outflow data become available. One more consequence of this picture is that the difference between the highest dense gas velocities in the acceleration region and dense gas velocities in the bulge should be even higher than the analogous difference for ionised gas.

M25 selected the ten galaxies because of their complex kinematics that were difficult to reproduce with standard modelling techniques, allowing them to showcase the power of their method. Our models, on the other hand, are spherically symmetric and we did not attempt to model the complex dynamics of the actual outflows, only the radial velocity trends. This simplification means that in principle, our results can be applied to any galaxy with a significant gaseous bulge component. However, we expect that if the outflow is smooth as it expands through the bulge, there should be less condensation of cold clumps, so the acceleration beyond the bulge should not be as rapid as in NGC 2992, NGC 5643, or NGC 5728. If it were discovered that galaxies with kinematically simpler outflows do not exhibit acceleration at all, or have significantly different radial velocity patterns from the ones investigated by M25, it would signify an interesting departure from the simple AGN wind feedback model and would warrant close investigation using 3D hydrodynamical simulations.

The spherically symmetric approximation of our model is also at odds with the conical outflow geometry adopted in M25. The 120-degree opening angle of the cone means the bi-conical outflow subtends a total solid angle $\Omega_{\rm cone} = 2\pi$, i.e. half of the full solid angle. Supposing that the whole of the AGN wind energy is injected into this solid angle, the effective luminosity experienced by the outflow would be twice the value adopted in our simulations. To get the correct outflow velocity, the gas fraction should be doubled as well (see Sect. \ref{sec:present_day_agn}). Since the volume of the conical sector is also 1/2 of the volume of the full spherical bulge, the total outflowing mass is the same. The total gas mass of the bulge is, however, at least twice the value obtained in the spherically symmetric approximation, because the gas in the region outside the outflow bi-cone must be at least as dense as in the bi-cone to confine it.

\subsection{Outflow driving mechanism} \label{sec:momentum_criticism}

M25 interpret the observed acceleration of AGN outflows as evidence of transition from momentum-driven to energy-driven flow. Within this paradigm \citep[see][for a detailed account]{King2003ApJ, King2010MNRASa, King2015ARAA}, the shocked AGN wind cools efficiently via IC scattering with the photons of the AGN radiation field inside a critical radius $R_{\rm C}$. Most of the wind energy is radiated away, and only the momentum is transferred to the surrounding ISM, leading to a relatively weak outflow. Outside $R_{\rm C}$, a significant fraction of the shocked wind energy contributes to pushing the ISM, leading to an energy-driven flow as described in Sect. \ref{sec:method}. Although this explanation seems qualitatively attractive, we highlight two shortcomings below.

The expected transition radius $R_{\rm C}$ can be estimated by comparing the IC cooling timescale and the momentum-driven flow timescale. Following the expressions given in \citet{King2003ApJ}, but allowing for an AGN luminosity that is a fraction, $l,$ of the Eddington luminosity and an arbitrary gas fraction, $f_{\rm g}$, we derive
\begin{equation}
    R_{\rm C} \approx 0.27 \left(M_{\rm BH,7} l f_{0.1}\right)^{1/2} \sigma_{150} \, {\rm kpc}
\end{equation}
using the same scaling as in Eq. \ref{eq:ven}. Inserting the SMBH masses and velocity dispersions of the observed galaxies, we find values of the cooling radius $0.07 < R_{\rm C}/{\rm kpc} < 1.9$, typically at least a factor of a few below the radius at which the acceleration begins. Considering that we expect the outflow to start accelerating somewhat before reaching $R_{\rm C}$ and accounting for the fact that, most likely, $f_{0.1} < 1$ and $l < 1$, the discrepancy becomes even higher. Only one galaxy, Centaurus A, has a cooling radius higher than the observed and modelled acceleration radius. However, as mentioned before, we suspect its velocity upturn may be spurious and actual outflow acceleration may begin at much higher radii.

The second issue relates to the absolute velocities of momentum-driven flows. We can estimate them using Eq. 14 of \citet{King2003ApJ} and making appropriate substitutions allowing for non-Eddington luminosities:
\begin{equation} \label{eq:vmom}
    v_{\rm mom} \simeq \left(\frac{4\pi G^2 M_{\rm BH} l}{2 f_{\rm g} \sigma^2 \kappa}\right)^{1/2} \simeq 85 \left(\frac{M_{\rm BH,7} l}{f_{0.1}}\right)^{1/2} \sigma_{150}^{-1} \,{\rm km \, s}^{-1}.
\end{equation}
Once we input the values of black hole mass and velocity dispersion, the expected momentum-driven outflow velocities fall in the range $45\, {\rm km \, s}^{-1}< v_{\rm mom} < 210\,{\rm km \, s}^{-1}$, which is a factor of $4-20$ below the actual outflow velocities observed in the central region. Allowing gas fractions lower than $0.1$ could bring velocities more in line with the observed values, but extreme adjustments would be required, i.e. the gas fractions would have to be $15-550$ times lower than the cosmological value $f_{\rm g} = 0.16$ and $2.5-10$ times lower than what we used in our models. This issue is exacerbated by the reduction of the expected velocities when we consider that usually $l<1$, i.e. AGNs are sub-Eddington. Furthermore, the calculation in \citet{King2003ApJ} does not account for the gravity of the galaxy; including that term leads to a further reduction in velocity, which can be approximated by replacing the term $(M_{\rm BH,7} l f_{0.1}^{-1})^{1/2}$ in Eq. \ref{eq:vmom} with $(M_{\rm BH,7} l f_{0.1}^{-1} - 1)^{1/2}$ \citep{King2010MNRASa}.

Finally, the very existence of a momentum-driven outflow region with a significant spatial extent is questionable. If we treat the shocked wind as a two-temperature plasma, the cooling time is governed not by the IC cooling rate of the electrons but by the energy equilibration between electrons and ions. This timescale can be several orders of magnitude longer than the IC cooling time. This leads to the formal cooling radius decreasing to $< 1$~pc so outflows are essentially always energy driven \citep{Faucher2012MNRASb}.

\section{Summary} \label{sec:sum}

We used a simple physical model, of large-scale adiabatic outflows driven by the wind energy input from a constant-luminosity AGN and expanding in an isothermal gaseous bulge, to fit the observed outflow radial velocity profiles of ten active galaxies. Our base model has only two free parameters, the gas fraction in the galaxy bulge and the bulge total mass. Six of the ten outflows can be fit very well by the base model, while one more can be fit by assuming a gas density distribution flatter than isothermal. The final three outflows also show an improved fit in the central regions when a flatter gas density distribution is assumed but accelerate faster than our model predicts in the outskirts. This may be a consequence of star formation within the outflowing material. The bulge masses required for our best-fit models agree very well with the observed SMBH--bulge mass relation in nine cases, with one (Centaurus A) requiring an under-massive bulge, although it is possible that the observed outflow acceleration is spurious and does not represent the outflow escaping the actual bulge. The required gas fractions also agree well with observed correlations of the gas fraction with the total stellar mass. The total gas mass in the bulge predicted in our models is always lower, usually by a significant factor, than the observationally estimated total gas mass of the host galaxies. Variations in other model parameters --- AGN luminosity and bulge velocity dispersion --- do not significantly impair the fits either and can be easily compensated for by adjusting the gas fraction and/or bulge mass.

These highly encouraging results provide strong validation for our proposed theoretical framework, especially given its conceptual simplicity. We suggest three observational priorities that will advance our understanding of outflow propagation in diverse environments and test our model. First, expanding outflow observations to encompass a broader sample of galaxies, including inactive ones, will illuminate the evolutionary diversity across different morphological types, gas reservoirs, and AGN luminosity histories. Second, extending radial profile measurements to greater galactic distances will test whether the observed acceleration trend persists or if there is a transition to the predicted deceleration phase. Third, similar observations of cold molecular gas outflows will reveal the complex interplay between shock cooling processes and turbulent ISM interactions that govern outflow evolution. These complementary observational advances promise to enhance our understanding of AGN outflow evolution and strengthen their utility as powerful diagnostic tools for investigating AGN luminosity histories.

\begin{acknowledgements} 
This research was funded by the Research Council Lithuania grant no. S-MIP-24-100. We thank Cosimo Marconcini for providing the outflow data used in our analysis and for very useful comments that helped improve the clarity of the paper.
\end{acknowledgements}

\bibliographystyle{mnras}
\bibliography{zubovas}

\label{lastpage}

\end{document}